\documentclass[preprint]{aastex63}
\usepackage{float}
\usepackage{wrapfig}
\usepackage{ifpdf}
\usepackage{microtype}
\DisableLigatures{encoding = *, family = * }
\received{August 31, 2020}
\revised{October 16, 2020}
\accepted{October 27, 2020}
\submitjournal{ApJ}

\shorttitle{No Stagnation Before the Heliopause at V1?}
\shortauthors{Cummings et al.}
\begin{document}

%%\title{\textbf{No Stagnation Region Before the Heliopause at Voyager 1? Inferences From New Voyager 2 Results}}
\title{No Stagnation Region Before the Heliopause at Voyager 1? Inferences From New Voyager 2 Results}

\correspondingauthor{A. C. Cummings}
\email{ace@srl.caltech.edu}

\author[0000-0002-3840-7696]{A. C. Cummings}
\affiliation{California Institute of Technology, Pasadena, CA 91125, USA}

\author{E. C. Stone}
\affiliation{California Institute of Technology, Pasadena, CA 91125, USA}

\author{J. D. Richardson}
\affiliation{Massachusetts Institute of Technology, Cambridge, MA 02139, USA}

\author{B. C. Heikkila}
\affiliation{NASA Goddard Space Flight Center, Greenbelt, MD 20771, USA}

\author{N. Lal}
\affiliation{NASA Goddard Space Flight Center, Greenbelt, MD 20771, USA}

\author{J. K\'{o}ta}
\affiliation{University of Arizona, Tucson, AZ 85721, USA}

%% \nocollaboration{2}



\begin{abstract}

We present anisotropy results for anomalous cosmic-ray (ACR) protons in the energy range $\sim$0.5-35 MeV from Cosmic Ray Subsytem (CRS) 
data collected during calibration 
roll maneuvers for the magnetometer instrument when Voyager 2 (V2) was in the inner heliosheath.
We use a new technique to derive for the first time the radial component of the anisotropy vector from CRS data. 
We find that the CRS-derived radial solar wind speeds,
when converted from the radial components of the anisotropy vectors via the Compton-Getting (C-G) effect,
generally agree with those similarly-derived speeds from the
Low-Energy Charged Particle experiment using 28-43 keV data. 
However, they often differ significantly from the radial solar wind speeds measured
directly by the Plasma Science (PLS) instrument. 
There are both periods when the C-G-derived radial solar wind speeds are significantly higher than those measured by 
PLS and times when they are significantly lower. 
The differences are not expected nor explained,
but it appears that after a few years in the heliosheath the V2 radial solar wind speeds 
derived from the C-G method  underestimate the true speeds as the spacecraft approaches the heliopause. 
We discuss the implications of this observation for the stagnation region reported along the Voyager 1 trajectory as it approached the 
heliopause  inferred using the C-G method.

\end{abstract}



\section{Introduction} \label{sec:intro}
%%%%%%%%%%%%%%%%%%%%%%%%%%%%%%%%%%%%%%%%%%%%%%%%%%%%%%%%%%%%%%%%%%
%%%%%%%%%%%%%%%%%%%%%%%%%%%%%%% %%%%%%%%%%%%%%%%%%%%%%%%%%%%%%%%%%%

The Voyager 1 (V1) Plasma Science (PLS)  instrument ceased functioning in 1980, and since that time
there have been no direct measurements of the plasma  properties from V1.
However, the Low-Energy Charged Particle (LECP) instruments on the two Voyagers
have been used to indirectly derive two components of the solar wind velocity vector by
using the well-known Compton-Getting (C-G) effect \citep{Fo70} in conjunction with
anisotropy observations at multi-keV energies \citep{KaDe98,KrRo11,DeKr12,KrDe13,RiDe14,RiBe20}.

In the case of V1, \citet{KrRo11} found that the C-G-derived radial solar wind speed gradually decreased to 0 km s$^{-1}$,
and even reported some small negative radial solar wind speeds, as the spacecraft
moved through the inner heliosheath and approached the heliopause.
They also found that the tangential component was trending towards zero across
the heliosheath as well.
It was thought that the solar wind had perhaps been deflected into the normal component
in order to turn and go down the tail of the heliosphere.

Up until March of 2011, the LECP experiment was not able to provide the normal component measurement, 
since it makes its anisotropy measurements by stepping a sensor in a single plane,
which is close to the R-T plane.\footnote{The RTN coordinate system is spacecraft 
centered with R pointing radially away from the Sun, T is parallel to the
solar equator and in the direction of the Sun's rotation, and N completes the right-handed system \citep{FrHa02}.}
However, in order to overcome that problem, the Voyager Project
arranged for the spacecraft to be re-oriented periodically, approximately every two months,  
for a few days at a time  by 70$^{\circ}$ so that the LECP scan plane
would include the direction of the normal component. 
The result was consistent with zero flow in the normal direction \citep{DeKr12}.

\citet{StCu11}  were also able to provide the normal component using data
from the Cosmic Ray Subsystem (CRS) experiment and using the same C-G technique with $\sim$0.5-35 MeV
proton data acquired during occasional rolls of the spacecraft designed to 
calibrate the magnetometers on the Voyager spacecraft.
Their results also showed that the normal component of the solar wind speed  was trending towards zero
across the heliosheath.

Thus, the idea of a stagnation region inside the heliopause along the V1 trajectory was born, where the solar wind seemed to come to
a stop before the heliopause was reached 
\citep{StCu11,BuNe12,OpDr12}. 
This was puzzling, since with the magnetic field frozen into the plasma, one would
expect the field magnitude to increase as V1 crossed the heliosheath
if all three components of the solar wind velocity were trending to zero.
It did not \citep{BuNe12}, and there was instead inferred a dramatic decline of the magnetic flux
at V1 \citep{RiBu13}, which was not expected.
It was conjectured that magnetic reconnection could be responsible \citep{OpDr12,RiBu13},
but \citet{DrSw17} concluded more study would be needed to see if that
mechanism was viable to explain the observations.
Another theory advanced was that solar cycle effects and/or heliopause instabilities are
playing  a role and it was shown that time-dependent simulations of the solar wind interaction with the very local 
interstellar medium (VLISM)  offered a plausible explanation of the zero and even
negative solar wind speeds inferred in the outer regions of
the inner heliosheath by the LECP instrument on V1 \citep{PoBo12,PoHe17}.

Voyager 2 (V2) has a working plasma instrument and there have been 
comparisons between the direct plasma speeds measured by PLS and the 
indirect derivations from the LECP experiment \citep{RiDe14,RiBe20}.
\citet{RiDe14}  examined data through the end of 2013 and 
found a period between 2009.3 and 2010.5, which they referred to as period A,  in which the  LECP
derivation of the radial solar wind speed was much higher than those measured directly
by PLS.
The authors speculated that  oxygen ions were getting into the LECP detector and contaminating the measurement.
\citet{RiBe20} extended observations out to just before the crossing of the heliopause by V2 on 5 November 2018.
They found other quasi-periodic variations in the C-G derived radial solar wind
speed that often did not match the steadier speeds from PLS.
And, the C-G speeds on average tended to trend lower than the direct measurements from PLS, 
which remained rather steady and only dropped towards zero very near the heliopause.

In this work, we present a new analysis of the CRS data that provides  
the radial component of the anisotropy vector, and hence a new C-G-derived estimate of the radial solar
wind speed.
We compare the CRS results with those of PLS and LECP and discover
a surprising result that casts doubt on the existence of the stagnation region reported from V1 observations.

\section{Observations} \label{sec:obs}
%%%%%%%%%%%%%%%%%%%%%%%%%%%%%%%%%%%%%%%%%%%%%%%%%%%%%%%%%%%%%%%%%%
%%%%%%%%%%%%%%%%%%%%%%%%%%%%%%%%%%%%%%%%%%%%%%%%%%%%%%%%%%%%%%%%%%

All observations in this work are from three instruments on the Voyager spacecraft: CRS \citep{StVo77}, 
PLS \citep{BrBe77}, and LECP \citep{KrAr77}. 
The CRS data were acquired from the Low-Energy Telescopes (LETs) during the ``magrol" maneuvers designed
to help the magnetometer team calibrate their instrument.
These magrols are a series of counter-clockwise rotations about the +R axis
when viewed from the Earth.
Prior to 2016, a single magrol maneuver consisted of 10 revolutions about the R axis.
However, due to power issues, there were no magrols in 2016 and 
the ones in 2017 and 2018 were limited to 1 or 2 revolutions. 
While the larger number of revolutions is helpful for statistical purposes,
we find we can still use the rolls with the smaller number of revolutions to fairly accurately
determine the anisotropy components.

In previous work using this type of analysis with CRS data \citep{StCu11,StCu17, CuSt19}, the authors fixed the R-component
of the observed anisotropy to be that expected from another instrument, which for V2 
was the radial solar wind speed from the PLS instrument converted to
a radial anisotropy component  in the spacecraft frame of reference using
the C-G effect \citep{Fo70}.
Since the roll is essentially about the R-axis, to determine the R-component from CRS observations alone
requires accurate knowledge of the responses of nearly-identical telescopes that
have radial components of their boresights with opposite signs.
As pointed out in the previous analysis papers, the actual value of the R-component has very little effect
on the derived T and N components of the anisotropy vector, which was the focus of those papers.
However, in this work we have determined the relative
response of the two telescopes employed  here by using an in-flight normalization  procedure
described below. 
Thus, we are able to determine all three components of the anisotropy vector from CRS data during magrols. 

%%%%%%%%%%%%%%%%%%%%%%%%%%%%%%% Figure 1 %%%%%%%%%%%%%%%%%%%%%%%%%%%%%%%%%%%
\begin{figure}
\begin{center}
\includegraphics*[trim= 0.7in 3in 0 1.1in,width=0.5\textwidth]{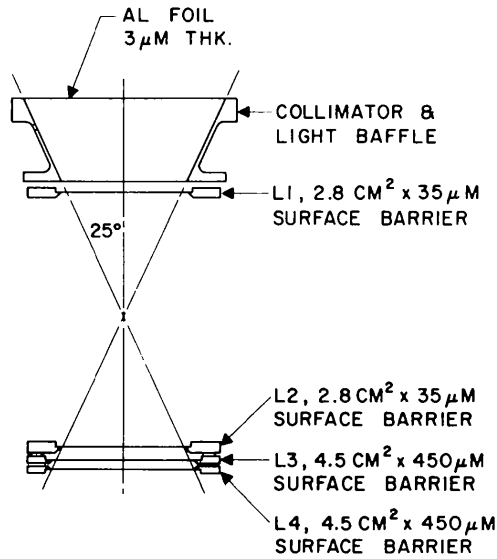}
\end{center}
\caption{Cross-section of a LET telescope (adapted from Stone et al., 1977).
Nominal areas and thicknesses of the detectors are shown in the figure. 
The 25$^{\circ}$ half-angle of the field of view shown is for the more typical analysis 
condition, in which the first two detectors must trigger. 
In the analysis used here during the magrols,
only the L1 detector is required to trigger.
The nominal distance between the entrance window and L1 is 1.524 cm and 
the nominal radius of the window is 1.7695 cm, resulting in
an 121$^{\circ}$ full acceptance angle for the particles.
%%The detectors are all silicon surface-barrier solid-state devices. 
}
%%\label{LET_telescope_Stone_et_al_1977}
\label{fig1}
\end{figure}
%%%%%%%%%%%%%%%%%%%%%%%%%%%%%%%%%%%%%%%%%%%%%%%%%%%%%%%%%%%%%%%%%%%%%%%%%%%%

The CRS data from the  magrols  consist of counting rates of particles triggering the first
detector in a  stack of four detectors that make up a LET telescope.
The cross-section of a LET is shown in Figure~\ref{fig1}.
There are four LETs on each of the Voyager spacecraft, referred to as LETs A, B, C, and D. 
They are identical in design, differing in practice only by small differences in
the characteristics of the nominally identical detectors and in the slightly different component spacings 
and positionings upon assembly.
The particles dominating the L1 rates are protons with $\sim$0.5-35 MeV and a median energy of $\sim$1.3 MeV  
that enter through the collimator opening at the top.
Additional  information about the L1 rates can be found  in Appendix~\ref{sec:app1}.

The four telescopes have their boresights arranged
in a quasi-orthogonal manner.
LETs A and C are mounted back-to-back and LETs B and D are
mounted with their boresights orthogonal to each other and to LETs A and C
\citep{StVo77}.

%%%%%%%%%%%%%%%%%%%%%%%%%%%%%%%  Figure 2 %%%%%%%%%%%%%%%%%%%%%%%%%%%%%%%%%%%
\begin{figure}
\begin{center}
\includegraphics*[trim= 0 4in  0 1.4in,width=0.7\textwidth]{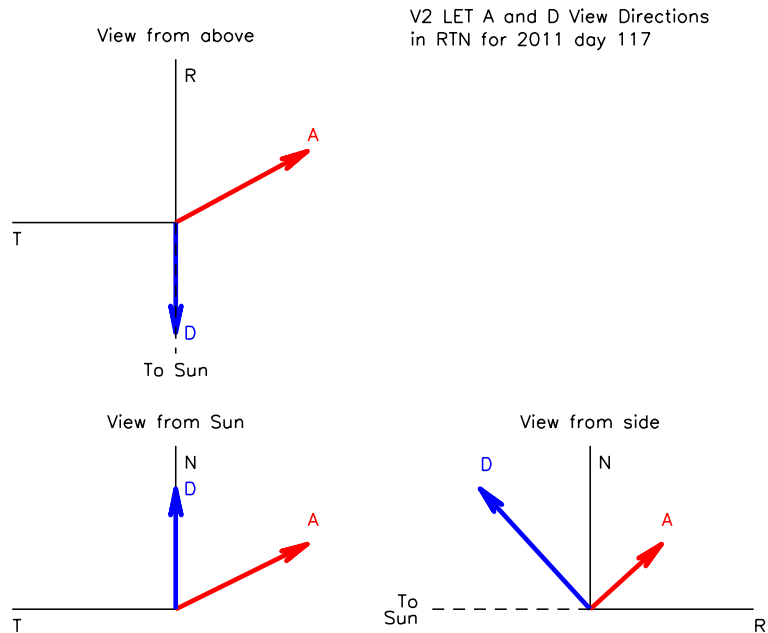}
\end{center}
\caption{
Three views showing the LET A and D telescope boresight unit vectors.
%%Boresights indicate the direction particles are coming from.
The arrows indicate the viewing directions of the central axis of the telescope boresights.
}
%%\label{LET_A_D_view_dir_plot_v2_2011_117}
\label{fig2}
\end{figure}
%%%%%%%%%%%%%%%%%%%%%%%%%%%%%%%%%%%%%%%%%%%%%%%%%%%%%%%%%%%%%%%%%%%%%

Soon after the V2 encounter with Jupiter in 1979, the LET B telescope ceased returning data and 
the L1 detector of LET C was judged  to have been implanted by sulfur and oxygen ions,
creating a thin layer, $\sim$2.9 $\mu$ Si equivalent thickness, that is insensitive or partially insensitive to particles passing through it.
This dead layer is believed to be slowly annealing \citep{Br85}.
To be sure this effect does not disturb the anisotropy results, and in particular
the new radial component analysis, we have elected to 
use only LETs A and D, omitting LET C data from the analysis. 

In Figure~\ref{fig2} we show the R, T, and N  components of the unit vectors representing 
the LET A and D boresights in three views.
The data are shown for 2011 day 117, which is representative
to within $\sim$2$^{\circ}$ for any time in the heliosheath when the
spacecraft is not undergoing  a maneuver.
Looking at the view-from-the-Sun diagram, the projection of the LET D boresight onto the N-T plane  is initially positioned at very near
0$^{\circ}$, with the N axis serving as the origin of the angle measurement.
The roll about the R axis is counter-clockwise in this view.
The roll will advance the two LET boresights in N-T angle and an anisotropy of the cosmic-ray intensity in the N-T plane 
will be revealed if it is large enough.

The other important view in  Figure~\ref{fig2} is the one from the side,
which shows that the boresight of LET D has a negative R-component,
whereas the boresight of  LET A has a positive R-component.
Thus, the R-component of the anisotropy vector can be determined if
the relative responses of LETs A and D are known accurately enough.

\begin{figure}
\begin{center}
\includegraphics*[trim= 0 6in 0 0,width=0.5\textwidth]{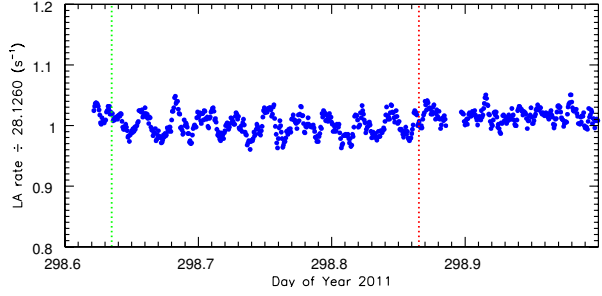}
\end{center}
\caption{Counting rate of mostly protons with $\sim$0.5-35 MeV from the 
LET A L1 detector 
(in anti-coincidence with the L4 detector (see Stone et al., 1977))
during ten spacecraft rolls on day 298 of 2011. 
The symbols are 5-point moving averages, where a point represents data collected
over a 48-second time period. 
The vertical green line marks the start of the roll and
the vertical red line marks the end of the roll.
}
%%\label{v2_la_magrol_fit63}
\label{fig3}
\end{figure}

In Figure~\ref{fig3} we show the counting rate from the LET A L1 detector 
(mostly protons with $\sim$0.5-35 MeV) during a typical 10-revolution roll maneuver on day 298 of 2011. 
The observed  periodic variation is consistent with a first-order anisotropy
with a period corresponding to the 2000 second duration of one 360$^{\circ}$
revolution of the spacecraft.

\begin{figure}
\begin{center}
\includegraphics*[trim= 0 95mm 0 36mm,width=0.5\textwidth]{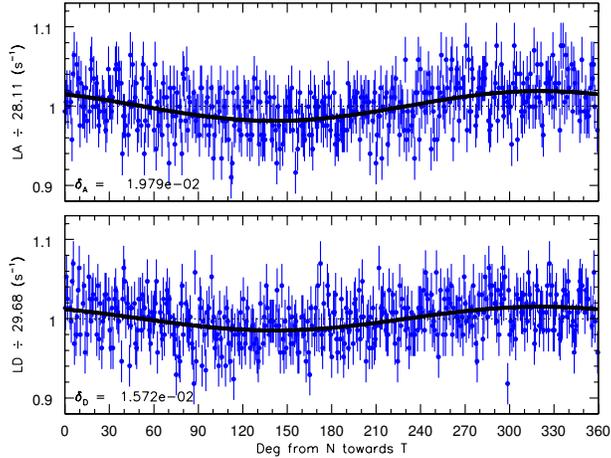}
\end{center}
\caption{(top panel) Counting rate from LET A L1 detector in the  L1$\overline{\textrm{L4}}$ command state from the magrol on day 298 of 2011 vs. angle 
from the N axis towards the T axis. 
The solid line represents the fit to a first-order anisotropy function
as described in the text.
The value of the anisotropy amplitude for the LET A data, $\delta_A$, 
derived from the fit is shown in the figure. 
(bottom panel) Same as top panel except for LET D.  
}
%%\label{v2_magrol_fitdelta_no_labels_063}
\label{fig4}
\end{figure}

In Figure~\ref{fig4} we show the data from both  the LET A and LET D L1 detectors
from the same roll as in Figure~\ref{fig3} but displayed as a function of roll angle rather
than time.
For all the rolls, LET A was in a command state such that the counting rate from L1 was one
that has L4 in anti-coincidence (see Figure~\ref{fig1}).
LET D was in that same command state for the data shown in Figure~\ref{fig4},
but that was not true for all magrols.
For the rolls before 2011/258 and for the roll on 2014/254, the anti-coincidence term was not present and the rate
was sampled only 1/16 of the time, whereas in the other command state the rate is
continuously measured.
We call the configuration in which LET D is in the statistically less favorable command state, configuration 1,
and when LET D was in the same command state as LET A the configuration is referred to as configuration 2.

%%%%%%%%%%%%%%%%%%%%%%%%%%%%%%% Figure 3 %%%%%%%%%%%%%%%%%%%%%%%%%%%%%%%%%%%
%%%%%%%%%%%%%%%%%%%%%%%%%%%%%%%%%%%%%%%%%%%%%%%%%%%%%%%%%%%%%%%%%%%%%%%%%%%%

%%%%%%%%%%%%%%%%%%%%%%%%%%%%%%%   Figure 4   %%%%%%%%%%%%%%%%%%%%%%%%%%%%%%%%%%%
%%%%%%%%%%%%%%%%%%%%%%%%%%%%%%%%%%%%%%%%%%%%%%%%%%%%%%%%%%%%%%%%%%%%%%%%%%%%%%

Also shown in  Figure~\ref{fig4} are the results of a simultaneous least-squares fit to a first-order anisotropy function
to the rates in each telescope: 
\begin{equation}
J_A = J_0 (1 + \vec{\delta} \cdot \vec{A}) + Abkg\label{eq:1}
\end{equation}
and  
\begin{equation}
J_D = k_D J_0 (1 + \vec{\delta} \cdot \vec{D}) + Dbkg\label{eq:2}
\end{equation}
where $\vec{A}$, for example, refers to the unit vector that
represents the boresight of LET A and $k_D$ represents
the response  normalization factor for LET D that  accounts for the slightly
non-identical geometry and response function to those of LET A.
The quantities $Abkg$ and $Dbkg$ represent a background rate that is assumed isotropic.
The method used to calculate these background rates is described in Appendix~\ref{sec:app2}.

The rates in Figure~\ref{fig4}
are plotted  versus  angle in the N-T plane
and thus the equations that are used in the fitting are written to accommodate
that situation.
For example, let $\beta_A$ represent the angle of the 
LET A boresight from N towards T in the N-T plane.
This angle advances 8.64$^{\circ}$ for each 48 s data point
and is known based on the original boresight vector, the start
time of the roll, and the roll rate. 
Let $\theta_A$ be the fixed angle of the 
LET A boresight vector from the R axis.
Then, the components of the LET A boresight vector are
$A_R = \cos \theta_A$, $A_T = \sin \theta_A \sin \beta_A$,
and $A_N = \sin \theta_A \cos \beta_A$.
Similar considerations apply to LET D.
The fit parameters are $J_0$ and the R, T, and N components
of the anisotropy vector, $\vec{\delta}$, with the normalization 
factor $k_D$ being determined as described below.

The R component of $\vec{\delta}$ depends critically on the factor $ k_D$.
In principle, this factor can be calculated by using a Monte Carlo simulation.
We have done such a calculation, which not only provides an estimate of $k_D$
but also provides the C-G factor \citep{Fo70} that will be needed to convert
between anisotropy vector components and solar wind speed components.
We describe obtaining the C-G factor first.

%%%%%%%%%%%%%%%%%%%%%%%%%%%%%%% Figure 5 %%%%%%%%%%%%%%%%%%%%%%%%%%%%%%%%%%%
\begin{figure}
\begin{center}
\includegraphics*[trim= 0 5.7in 0 1.4in,width=0.5\textwidth]{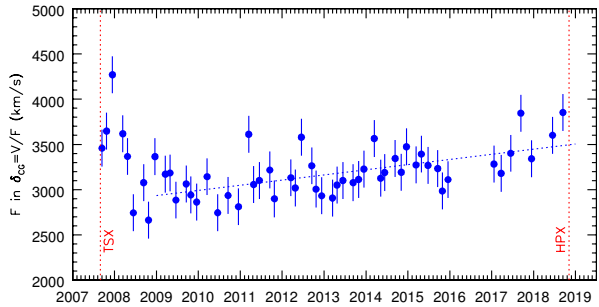}
\end{center}
\caption{
Compton-Getting factor for converting a solar wind velocity component to an anisotropy component as described in the text.
For reference, an E$^{-1.5}$ spectrum gives F = 3303 km s$^{-1}$ from the Monte Carlo simulation for LET A. 
The dotted line was used to estimate the uncertainty on each point.
}
%%\label{v2_KCG_time_with_sig}
\label{fig5}
\end{figure}
%%%%%%%%%%%%%%%%%%%%%%%%%%%%%%%%%%%%%%%%%%%%%%%%%%%%%%%

The C-G anisotropy, $\vec{\delta_{CG}}$, is given by
\begin{equation}
\vec{\delta_{CG}} = <(2-2\gamma)/v>\vec{V}\label{eq:3}
\end{equation}
where $v$ is particle speed, $\vec{V}$ is the
solar wind velocity, and $\gamma$ is the power-law index
in the differential energy spectrum, dJ/dE $\sim$ E$^{\gamma}$,
and the brackets $<>$ denote the average over the energy spectrum of the enclosed quantity.
We have fit the energy spectra data to a four-power-law function
for each roll day (or an adjacent roll day if the roll day had poor statistics) and then employed a Monte Carlo simulation using that function
from 0.4-40 MeV to select particles and input them on the top of the aluminum window of a telescope.
See Appendix~\ref{sec:app1} for an example of the resulting energy loss distribution in an L1 detector
and how it compares with the observed distribution. 
The Monte Carlo simulation routine uses the same proton range-energy formulation that is used in other parts of  the analysis of
CRS data \citep{Co81} and keeps track of whether energy losses in each detector exceed their threshold for triggering. 
One result of the simulation for the LET A telescope  is the reciprocal of the coefficient of $\vec{V}$ in equation~\ref{eq:3},
which we show in Figure~\ref{fig5}.
An uncertainty of 204 km s$^{-1}$ on each point was derived by assuming the linear function shown
by the dotted line represents the data from 2009 forward and by
adjusting the uncertainty on each point to give a reduced $\chi^2$ = 1.0. 

As mentioned, another result of the Monte Carlo simulation is the factor $k_D$, 
which is found by comparing the rates triggering the LET A and LET D 
L1 detectors in their appropriate command states for the same input energy spectrum,
which itself varies somewhat over time and/or distance in the heliosheath.  
We found this calculation resulted in $k_D$ factors that were too small
to make the converted radial solar wind speed, VR, agree with 
the PLS and LECP results when those two instruments agreed.
This is likely due to incomplete knowledge of the exact areas of the detectors and collimator and/or
spacings between telescope elements and their positionings that determine the effective geometrical factor
of each telescope.
The variation of the calculated $k_D$ factors over the 55 magrols in the
heliosheath was small, typically much less than 1\%.
The variation is due to a slight dependence of the effective
geometrical factor on the shape of the energy spectrum.

\begin{figure}
\begin{center}
\includegraphics*[trim= 0 6.5in 0 1.4in,width=0.5\textwidth]{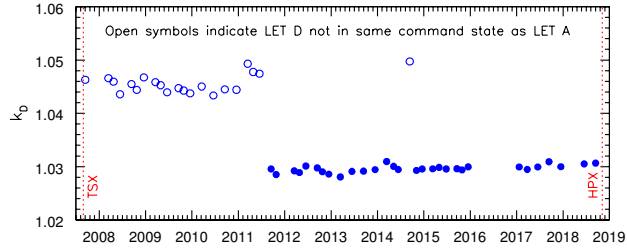}
\end{center}
\caption{
Normalization factor $k_D$ as a function of time that was used in fitting Equation~\ref{eq:2}
as described in the text.
The solid symbols indicate rolls when LET A and D were in the same command state
(configuration 2). The open symbols denote the $k_D$ for configuration 1.
}
%%\label{v2_kD_time}
\label{fig6}
\end{figure}

Our approach to this problem was to first do a series of fits in which 
the R-component of the anisotropy vector was fixed to be the value expected from
the radial solar wind speed at the time of the roll and was obtained from the PLS measurements.
We then picked normalization rolls, when PLS and LECP results for VR were
in agreement; three when LET D was in the same
command state as LET A (configuration 2; both in the L1$\overline{\textrm{L4}}$ mode)  
and three when LET D was in the L1 mode (configuration 1).
The agreement between PLS and LECP for VR was determined from Figure 6 of
\citet{RiDe14}.
The following rolls were used as normalizers for $k_D$ for configuration 1 (year/day of start of roll):
2008/73, 2008/114, 2008/353.
For configuration 2, the normalization rolls were on 2013/73, 2013/164, and 2013/302.
In each case the average of the $k_D$ from the simulation was compared to that
from the $k_D$ that would result in a VR that would agree with PLS results (and LECP results).
An average factor was determined that could be applied to the $k_D$ from the simulation
to get the $k_D$ as a function of the rolls to use in a new set of fits,
in which the R-component of the anisotropy vector was now a fit parameter.
The two factors were 1.04294 for configuration 1 and 1.03106 for configuration 2.
The resulting $k_D$ is shown in Figure~\ref{fig6}.
 
%%%%%%%%%%%%%%%%%%%%%%%%%%%%%%%   Figure 6     %%%%%%%%%%%%%%%%%%%%%%%%%%%%%%%%%%%
%%%%%%%%%%%%%%%%%%%%%%%%%%%%%%%%%%%%%%%%%%%%%%%%%%%%%%%%%%%%%%%%%%%%%%%%%%%%%%%%%%%%%%

%%%%%%%%%%%%%%%%%%%%%%%%%%%%%%% Figure 7 %%%%%%%%%%%%%%%%%%%%%%%%%%%%%%%%%%%
\begin{figure}
\begin{center}
\includegraphics*[trim= 0 2in 0 1.25in,width=0.5\textwidth]{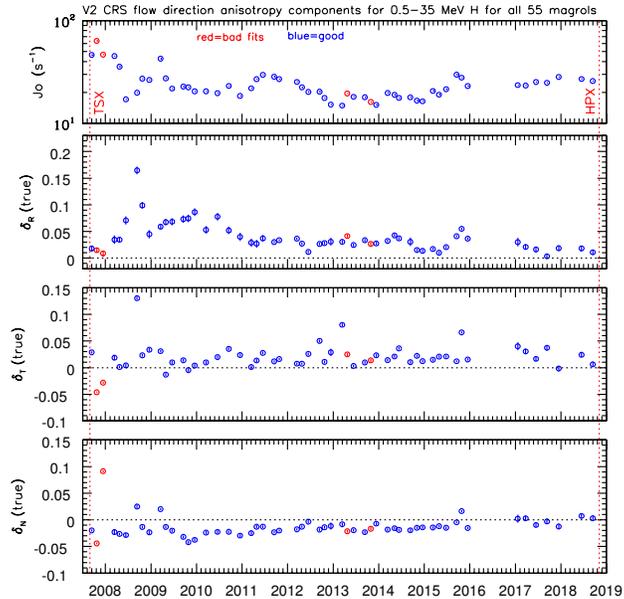}
\end{center}
\caption{(top panel) Parameter $J_0$ from fits described in the text
for 55 magrols when V2 was in the inner heliosheath. 
The red points represent fits that were judged to be not good by a $\chi^2$ test.
The 51 blue points represent rolls that resulted in good fits.
(second panel) Values of the R-component of the anisotropy amplitude from the fits.
The designation ``true" is described in the text.
(third panel) Same as second panel except for the T-component of the anisotropy amplitude. 
(bottom panel) Same as third  panel except for the N-component of the anisotropy amplitude. 
The vertical  lines mark the times of the termination shock crossing (TSX)  and the heliopause crossing (HPX).   
}
%%\label{v2_CRS_delta_components_time}
\label{fig7}
\end{figure}
%%%%%%%%%%%%%%%%%%%%%%%%%%%%%%%%%%%%%%%%%%%%%%%%%%%%%%%%%%%%%%%%%%%%%%%%%%%%%%%%%%%%%%

We note that the counting rates that result from an anisotropy in the intensity of particles
depends on the geometry of the telescope.
The effect of the 121$^{\circ}$ field of view was addressed in a Space Radiation Lab Internal Report \#84 (1981) by N. Gehrels and D. Chenette
(available upon request).
From the internal report, it can be inferred that the reduction factor of the amplitude of  the $\vec{\delta}$ that exists in space
is the ratio of two integrals
[$2\pi\int_{0}^{\pi}A(\theta)sin\theta cos\theta d\theta]/[2\pi\int_{0}^{\pi}A(\theta)sin\theta d\theta$]
where $A(\theta)$ is the overlap area of two co-aligned, circular disks as a function of angle from
the boresight axis and is given by Equation 10 of \citet{Su71}.
According to the internal report, for the nominal LET telescope geometry, the reduction  factor would be 3.8744/4.5964 = 0.8429. 
For the actual geometry of V2 LET A, we calculate the reduction factor to be 3.9573/4.7203 = 0.8384, which is the
value used here.

In addition, we note that the $\vec{\delta}$ from the fits represents the direction from which the
particles are coming.
Thus, to represent the results for the actual flow direction  and to correct
for the 121$^{\circ}$ field of view of
the telescopes, 
we have multiplied the components from the fits by a factor
of -1/0.8384 = -1.193
and labeled them as ``true" in Figure~\ref{fig7}.
Also, the radial component has been increased in $\delta$ by the equivalent of 15 km s$^{-1}$, the
spacecraft radial speed, to
move the results out of the spacecraft frame of reference into the Sun frame.
The figure shows the results for $J_0$ 
and the true R, T, and N components of 
$\vec{\delta}$ for all 55 magrols that occurred between
the V2 crossing of the TS on 30 August 2007 and its crossing
of the heliopause on 5 November 2018.
The four rolls that were judged not to yield good fits by the Q statistic test 
are shown as red symbols in Figure~\ref{fig7}.\footnote{The Q statistic 
is the complement of the chi-square probability function (\citet{PrTe92}, equation 6.2.19)
and the fit was judged to be good if its value was within the range 0.05 to 0.999.} 
The results for the 51 good-fit rolls are shown in tabular form  in Appendix~\ref{sec:app3}.
Note that in the tables we have numbered the rolls using our own system in which roll 1 was on day 123 of 2001.

\section{Radial Component of Solar Wind Velocity} \label{sec:VR}
%%%%%%%%%%%%%%%%%%%%%%%%%%%%%%% %%%%%%%%%%%%%%%%%%%%%%%%%%%%%%%%%%%
%%%%%%%%%%%%%%%%%%%%%%%%%%%%%%% %%%%%%%%%%%%%%%%%%%%%%%%%%%%%%%%%%%

\begin{figure}
\begin{center}
\includegraphics*[trim= 0 2in 0 1.25in,width=0.6\textwidth]{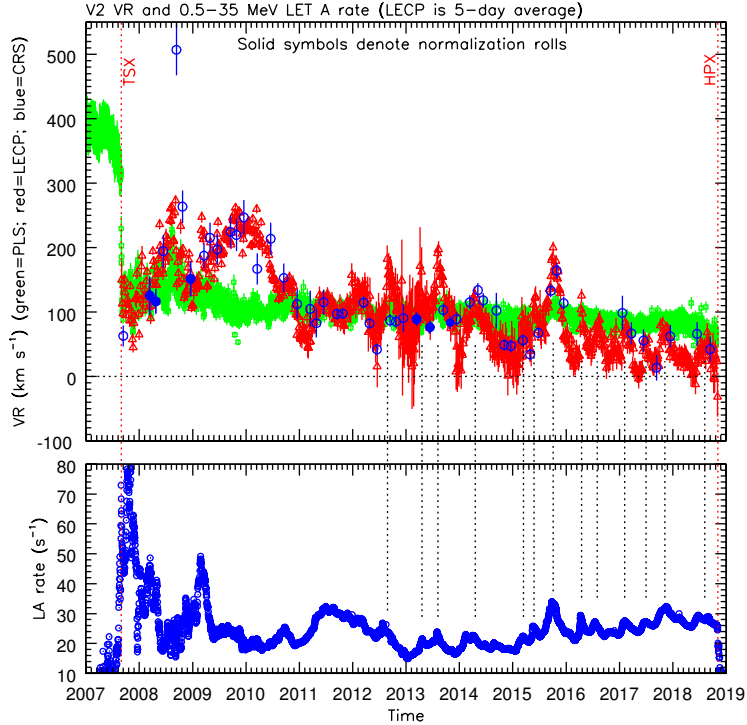}
\end{center}
\caption{
(top panel) Radial component of the solar wind velocity from CRS (blue), PLS (green), and LECP (red).
LECP (5-day averages) and CRS data  (for all 51 good-fit rolls)  are derived values based on the C-G effect. 
PLS data are daily averages and are direct measurements.
The six periods used to normalize the LET D to LET A rates are shown as the solid circles. 
The LECP data are from a combination of data presented in \citet{RiDe14} and \citet{RiBe20}.
(bottom panel) Daily-averaged LET A L1$\overline{\textrm{L4}}$ rate ($\sim$0.5-35 MeV protons).
Starting in mid-2012, most of the increases in this rate align with increases in the LECP VR values.
The dotted vertical lines are drawn to illustrate the correlation.
These LET A L1$\overline{\textrm{L4}}$  rate increases are indicators of pressure pulses
discussed in \citet{RiBe18}.
During the quieter times between the pressure pulses, the LECP and CRS VR values are significantly
below the actual measured VR values from PLS.
}
%%\label{v2_VR_AL1notL4_2007_2018}
\label{fig8}
\end{figure}

The derivation of the radial component of the anisotropy vector
during magrols for $\sim$0.5-35 MeV protons from CRS observations is a new development.
As noted earlier, this radial component can be converted to the 
convective radial solar wind speed, VR, using the C-G method \citep{Fo70}.
The observations for VR from CRS, along with those from PLS and LECP,
are shown in  Figure~\ref{fig8}.
Except for one CRS very high value in 2008 (from magrol 46 on day 255 of 2008),
which represents an unusual streaming event that will be the subject of a future study,
there is remarkable agreement between the CRS and LECP results, including for 
the previously discounted 2009.3-2010.5 period (period A in \citet{RiDe14}).
Yet, the CRS and LECP inferred speeds are often different from
the directly measured speeds from PLS.
We note that because of the agreement between CRS and LECP during 2009.3-2010.5, 
it now appears that oxygen ions were not contaminating the LECP data
during that period.

%%%%%%%%%%%%%%%%%%%%%%%%%%%%%%%   Figure 8     %%%%%%%%%%%%%%%%%%%%%%%%%%%%%%%%%%%
%%%%%%%%%%%%%%%%%%%%%%%%%%%%%%%%%%%%%%%%%%%%%%%%%%%%%%%%%%%%%%%%%%%%%%%%%%%

Just after the termination shock crossing, for a period of  about 1 year, all three instruments gave
approximately the same results for VR.
Then, there is the afore-mentoned period from $\sim$2009.3-2010.5,  
when the CRS and LECP values were much higher
than those directly measured by PLS, by as
much as a factor of $\sim$2.5.
Starting soon after that, the inferred radial solar wind speeds 
from  LECP appeared to oscillate, at first on average near the PLS measurements, which were not
oscillating, and then they trended down to below the PLS measurements as
V2 moved closer to the heliopause.

%%%%%%%%%%%%%%%%%%%%%%%%%%%%%%   Figure 9    %%%%%%%%%%%%%%%%%%%%%%%%%%%%%%%%%%%
\begin{figure}
\begin{center}
\includegraphics*[trim= 0 2in 0 1.25in,width=0.6\textwidth]{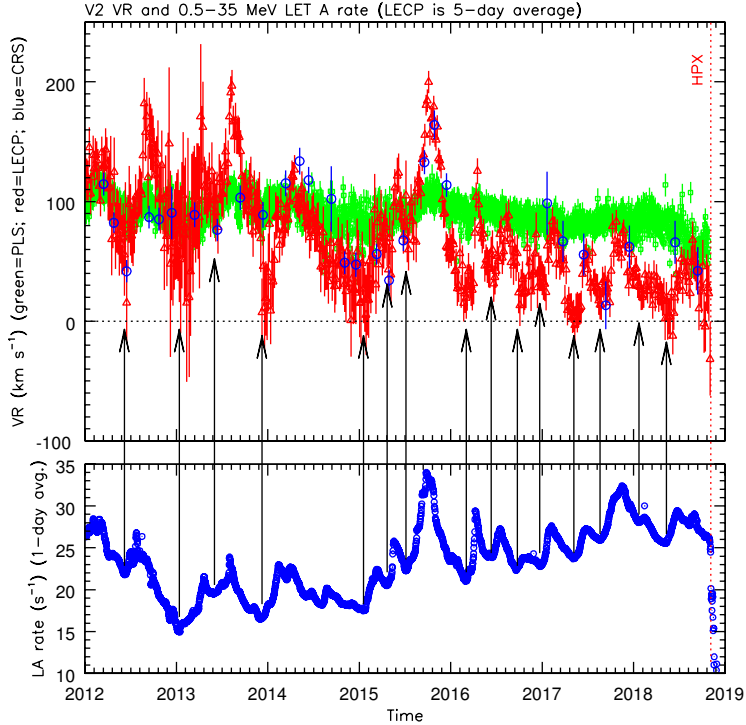}
\end{center}
\caption{
Same as Figure~\ref{fig8} except for time period shown and the  vertical scales have been changed.
The arrows show the correspondence of the local minima in the LET A L1$\overline{\textrm{L4}}$ rate and the local
minima in the LECP inferred radial solar wind speeds.  
}
%%\label{v2_VR_AL1notL4_2012_2018}
\label{fig9}
\end{figure}
%%%%%%%%%%%%%%%%%%%%%%%%%%%%%%%%%%%%%%%%%%%%%%%%%%%%%%%%%%%%%%%%%%%%%%%%%%

\citet{RiBe18} describe possible pressure waves in the heliosheath at V2. 
The counting rate of $\sim$0.5-35 MeV protons track these pressure changes and 
this rate is shown in the bottom panel of Figure~\ref{fig8}.
We note that after $\sim$mid-2012 there appears to be a good correlation of local maxima in
this rate with local maxima in the VR
from  LECP and CRS. 

If one were to consider only the lower pressure regions,
the LECP data would show VR values near zero as
the heliopause was approached, similar to that
which was observed at V1.
This is shown more clearly in Figure~\ref{fig9},
where the arrows show the correspondence of the local minima in the
$\sim$0.5-35 MeV rate with the local minima in the LECP and CRS VR speeds.
However, the PLS VR values did not approach zero but were near
80 km s$^{-1}$ during that time.
This finding then calls into question the stagnation region reported as V1 approached the heliopause,
which was based on the V1 LECP inferences of VR using the C-G effect that showed
VR trending towards zero as the heliopause was approached
\citep{KrRo11}.

We do not understand why there is such good agreement, in general, between CRS
and LECP VR results during times when PLS is different. 
Ordinarily, if the C-G method gives the same
result at two different energies, not to mention two
widely different energy bands as we have here, it would be confirmation
that the correct solar wind speed had been deduced.
But, that is apparently not the case with these measurements.

If the differences are attributed to the presence of a diffusive particle flow, the agreement between CRS and LECP for VR would imply an
R component of the diffusive particle flow vector that
differs at 28-43 keV versus that at $\sim$0.5-35 MeV by
the ratio of their C-G factors, which would be 3303/513 = 6.4 for
an E$^{-1.5}$ spectrum.
The diffusive anisotropy depends on particle speed, $v$, the diffusion tensor, $K$,
and the gradient of the number density per unit energy, $U$ \citep{Fo70} according to:
\begin{equation}
\vec{\delta_{\rm{diff}}} = -(3/v)  K \cdot \bigtriangledown U/U\label{eq:4}
\end{equation}
Near the heliopause, where the C-G-derived radial speeds are below those of PLS,
the diffusive flow would need to be directed inwards to explain the observations.

The Monte Carlo simulations revealed that the typical median energy
for the $\sim$0.5-35 MeV interval is $\sim$1.3 MeV (see Appendix~\ref{sec:app1}). 
So, the factor of $\sim$6.4 described above would imply an $\sim$$E^{-0.5}$ energy dependence to 
the product of the factors in Equation~\ref{eq:4}.
The study of this phenomenon is deferred to the future.

\section{Summary} \label{sec:sum}
%%%%%%%%%%%%%%%%%%%%%%%%%%%%%%% %%%%%%%%%%%%%%%%%%%%%%%%%%%%%%%%%%%
%%%%%%%%%%%%%%%%%%%%%%%%%%%%%%% %%%%%%%%%%%%%%%%%%%%%%%%%%%%%%%%%%%

We have used data from CRS acquired during occasional rolls of the V2 spacecraft 
when V2 was in the inner heliosheath
and deduced the radial, tangential, and normal components of the anisotropy flow vector
for $\sim$0.5-35 MeV protons.
The measurement of the radial component of the anisotropy vector is a new development for CRS data.
We have converted those radial components into radial solar wind speeds
using the C-G method and compared with results from LECP, derived in a similar
way, and also with the direct measurements from PLS.

We find that there is more than one aspect to the comparisons.
There is an initial period of $\sim$1 year after the termination shock crossing 
when all three instruments give approximately the same results,
except for one CRS result in 2008 that deserves a separate study.
That is followed by a period of $\sim$1.2 years, the period A in \citet{RiDe14}, when the C-G-derived results from LECP and CRS agree
with each other,
but are much higher, up to a factor of $\sim$2.5 times higher, than the direct measurements from PLS.
We do not offer an explanation for this difference.
However, we note that
\citet{RiDe14} speculated that the LECP measurements during period A  may have been contaminated
by oxygen ions.
The new inferences of VR from CRS are key to ruling out this explanation 
and point to an unexplained phenomenon at work.

After $\sim$mid-2012, beginning approximately five years after the termination shock crossing, 
there is a remarkable correlation between pressure pulses,
characterized by changes in the plasma density and particle intensities, 
and variations in the CRS and LECP C-G-derived radial solar wind speeds.
Often these speeds do not agree with the direct measurements from PLS.
In the low pressure regions the LECP VR values trend towards zero,
similar to the phenomenon that was seen at V1.
However, PLS on V2 did not observe a real trend of VR towards zero across the heliosheath.
Thus, the question is raised about whether the trend  in VR inferred by LECP at V1 was a real trend of the
radial solar wind speed or not.

While it is plausible that the solar wind might stagnate somewhere in the outer heliosphere,  the fact that
the V1 magnetic field strength did not increase proportionally argues against it happening along the V1 trajectory.
The assertion that there is a stagnation region was made by instruments on V1
that used the C-G method to infer the solar wind speeds.
There is no working plasma instrument on V1 with which to make direct measurements.
Now we have the situation on V2 where the C-G method used by
two instruments using very different energy ranges are getting the same VR, 
but they are not always in agreement with the direct
measurements from the working plasma instrument on V2.
In particular, in the vicinity of the heliopause at V2, even the average of
the oscillatory VR inferred from LECP observations using the C-G method is significantly
below the VR from PLS and trends downwards with time.
Also, the values during the minima of the C-G VR oscillations are near zero km s$^{-1}$,
while PLS is measuring radial speeds near 80 km s$^{-1}$.

Assuming the PLS speeds are correct, some phenomenon is interfering 
with the C-G method's ability to give the correct radial speeds, at least at V2.
And the interference is in such a way that CRS at $\sim$0.5-35 MeV and LECP at 28-43 keV
are getting the same incorrect answer.
If this same phenomenon is operating along V1's trajectory through the inner heliosheath,
then the implication is that there is no stagnation region before the heliopause at V1.

The trends toward zero speeds across the heliosheath at both V1 and V2 for the other two components, $\delta_T$ and $\delta_N$,  have been
addressed previously \citep{StCu11,StCu17,CuSt19} and the results suggest that a diffusive flow of ACRs
is responsible, at least in the case of $\delta_T$. 
Updates of those studies, based on the new techniques outlined in this work for CRS, are planned for the future.

\acknowledgments
%%%%%%%%%%%%%%%%%%%%%%%%%%%%%%%%%%%%%%%%%%%%%%%%%%%%%%%%%%%%%%%%%
%%%%%%%%%%%%%%%%%%%%%%%%%%%%%%% %%%%%%%%%%%%%%%%%%%%%%%%%%%%%%%%%%%

ACC, ECS, NL, and BCH  acknowledge Voyager data analysis support from NASA award number NNN12AA01C.
JDR was supported under NASA contract 959203 from the Jet Propulsion Laboratory to the Massachusetts 
Institute of Technology.
ACC and JDR also acknowledge support by NASA grant 
18-DRIVE\_2-0029, Our Heliospheric Shield, 80NSSC20K0603.

%%\clearpage
\appendix
%%%%%%%%%%%%%%%%%%%%%%%%%%%%%%%%%%%%%%%%%%%%%%%%%%%%%%%%%%%%%%%%%%
%%%%%%%%%%%%%%%%%%%%%%%%%%%%%%% %%%%%%%%%%%%%%%%%%%%%%%%%%%%%%%%%%%

\section{Characteristics of the ~0.5-35 MeV Rate} \label{sec:app1} 
%%%%%%%%%%%%%%%%%%%%%%%%%%%%%%%%%%%%%%%%%%%%%%%%%%%%%%%%%%%%%%%%%%%
%%%%%%%%%%%%%%%%%%%%%%%%%%%%%%% %%%%%%%%%%%%%%%%%%%%%%%%%%%%%%%%%%%

Several quantities of interest to this investigation rely on a Monte Carlo code
written to simulate the response of the L1 detectors to an input energy spectrum.
In this Appendix, we show some of the results from the Monte Carlo program.

The code uses a range vs. energy relationship for protons in Si that was used
in other parts of the analysis of CRS data (see, e.g., \citet{Co81} and \citet{CoSt84}).
A proton with an energy randomly selected from a differential energy spectrum, in the range
from 0.4-40 MeV, and with a random trajectory selected from an isotropic angular distribution, 
is input at a random position on the front Al window (see Figure~\ref{fig1}).
The particle is followed along a straight-line trajectory and intersections with any
detectors are noted. 
Energy loss in each detector layer is calculated and compared to the threshold energy
for triggering that detector and that information is used in determining how many
successful triggers occurred for various coincidence conditions.

The model of the LET telescope used in the simulation includes information on vertical detector spacings noted during their assembly,
as well as thicknesses and areas of the detectors determined from pre-launch measurements. 
The average thicknesses of the two front detectors, L1 and L2, were measured prior to launch in five concentric rings, each 2 mm in width, 
using radioactive sources, and those data are incorporated into the simulations. 

%%%%%%%%%%%%%%%%%%%%%%%%%%%%%%% Figure 10 %%%%%%%%%%%%%%%%%%%%%%%%%%%%%%%%%%%
\begin{figure}
\begin{center}
\includegraphics*[trim= 0 2.7in 0 1.2in,width=0.7\textwidth]{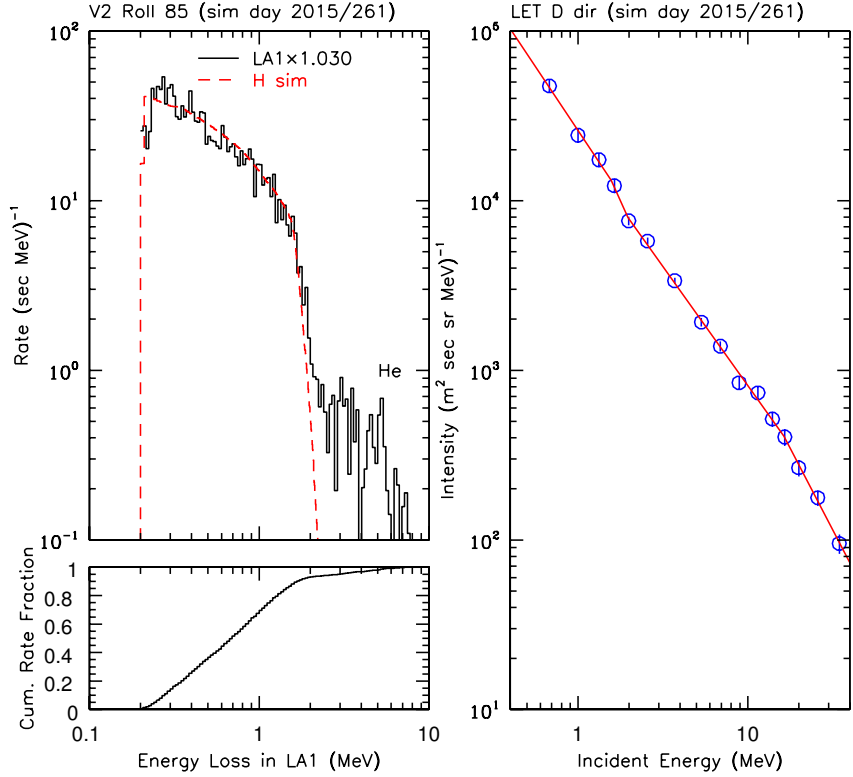}
\end{center}
\caption{(top right) Incident energy spectrum of protons appropriate for the LET D direction for day 261 of 2015.
The observations (open circles) are from CRS. The red line is a fit of the data to  a four-power-law
function that is used as input to the Monte Carlo simulation program described in the text.
(top left) Energy-loss distribution in LET A L1 detector for the  L1$\overline{\textrm{L4}}$ coincidence condition from the simulation (dashed red line) along with the observed 
energy-loss distribution (histogram). 
Above $\sim$2 MeV the observed distribution is dominated by
He ions.
(lower left) Cumulative fraction of the rate vs. energy loss in LET A L1. 
}
%%\label{v2la1_energy_dist_2015_261}
\label{fig10}
\end{figure}
%%%%%%%%%%%%%%%%%%%%%%%%%%%%%%%%%%%%%%%%%%%%%%%%%%%%%%%%%%%%%%%%%%%%%%%%%%%%%%%

In Figure~\ref{fig10} we show both an input energy spectrum 
and the resulting energy loss distribution in LET A L1 for the coincidence condition L1$\overline{\textrm{L4}}$,
which was the only coincidence condition for LET A used in this work.
The incident energy spectrum was measured by CRS and is for 2015/261, which is the day after the
actual magrol (number 85).
It is often necessary for statistical reasons to use a day adjacent to the actual day of the magrol
for the simulations.
The input to the Monte Carlo program is the four-power-law fit to the data shown in the right panel. 

The four lowest-energy data points of the input energy spectrum are from analysis of the observed energy-loss measurements 
shown in the left panel of the figure, after corrections for background.
These four points have been corrected to be as if they were from  LET D  by a factor determined
by the relative counting rates of LET A L1 and LET D L1.
This was necessary because the LET A multi-detector data, used for the next five highest-energy points,
have very poor statistics, and LET D was used for these.
The remaining points are from the High-Energy Telescope ( HET) 1.
More information about the multi-detector data analysis can be found in \citet{CuSt16}.

Since the measured energy spectrum and the simulation are appropriate for the LET D direction, the observed energy loss distribution
from LET A L1$\overline{\textrm{L4}}$, shown in the top left panel, has been multiplied by a factor to account for that difference.
The factor is taken from Figure~\ref{fig6} and the procedure describing how it was obtained
is described in the text.

The simulated energy-loss distribution and the observed energy-loss distribution are in good agreement below $\sim$2 MeV. 
Above that energy, the observed energy-loss distribution is dominated by He ions.
(See \citet{StCu03} for simulations using He and heavier ions.)
Based on the lower left panel of Figure~\ref{fig10}, $\sim$95\% of the rate responds to protons.
The He ions are anomalous cosmic rays and would be expected to have
similar anisotropy characteristics to those of the protons, so they are not treated as 
background.
Rather, the anisotropy study in this work is regarded as pertaining to
a population of particles dominated by protons with $\sim$0.5-35 MeV.

%%%%%%%%%%%%%%%%%%%%%%%%%%%%%%% Figure 11 %%%%%%%%%%%%%%%%%%%%%%%%%%%%%%%%%%%
\begin{figure}
\begin{center}
\includegraphics*[trim= 0 1.7in 0 1.2in,width=0.5\textwidth]{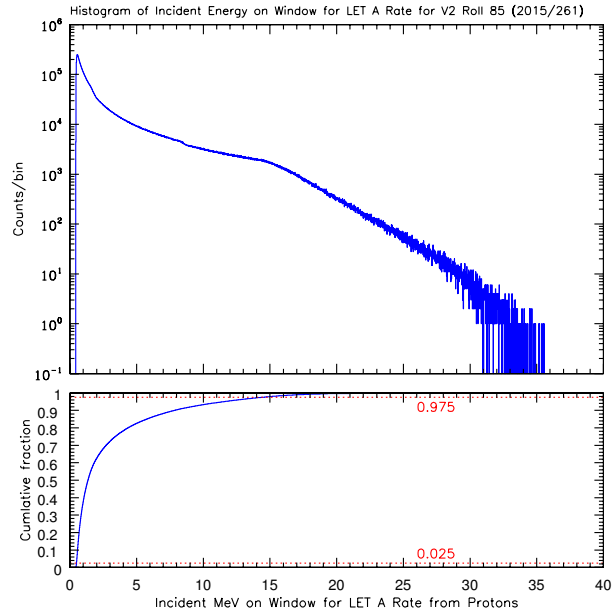}
\end{center}
\caption{(top) Histogram of the incident energies for the same simulation used to produce the results
shown in Figure~\ref{fig10}.
Incident protons above $\sim$35 MeV at the extreme allowable angle to the normal
deposit too little energy in L1 to trigger the threshold of the L1 detector.
Incident protons below $\sim$0.5 MeV are also unable to trigger the threshold. 
(bottom) Cumulative counts of the distribution shown in the top panel vs. incident energy
expressed as a fraction of the total.
The two horizontal dotted lines indicate that 95\% of the distribution is from $\sim$0.5-14 MeV.
}
%%\label{v2_L1notL4_inc_energy_histo_85}
\label{fig11}
\end{figure}
%%%%%%%%%%%%%%%%%%%%%%%%%%%%%%%%%%%%%%%%%%%%%%%%%%%%%%%%%%%%%%%%%%%%%%%%%%%%%%%%%%%%%%

%%%%%%%%%%%%%%%%%%%%%%%%%%%%%%% Figure 12 %%%%%%%%%%%%%%%%%%%%%%%%%%%%%%%%%%%
\begin{figure}
\begin{center}
\includegraphics*[trim= 0 4.7in 0 1.25in,width=0.5\textwidth]{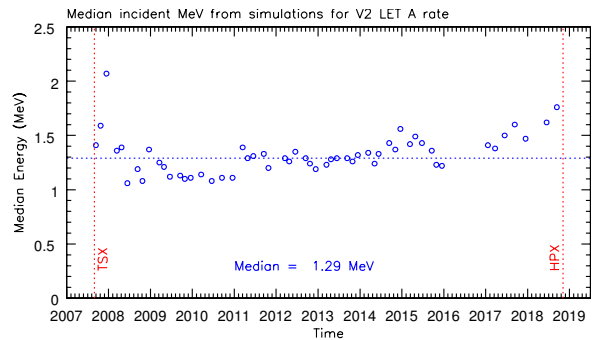}
\end{center}
\caption{Median incident energies from the distributions like those in Figure~\ref{fig11} for all 55 magrols vs. time.
}
%%\label{v2_Einc_median_time}
\label{fig12}
\end{figure}
%%%%%%%%%%%%%%%%%%%%%%%%%%%%%%%%%%%%%%%%%%%%%%%%%%%%%%%%%%%%%%%%%%%%%%%%%%%

In Figure~\ref{fig11} we show the histogram of incident energies
that resulted in LET A L1$\overline{\textrm{L4}}$ triggers for the same simulation 
used in Figure~\ref{fig10}.
It is apparent why the energy interval ascribed to the L1 rates is $\sim$0.5-35 MeV.
However, as shown in the lower panel, $\sim$95\% of the rate is due to
protons with $\sim$0.5-14 MeV.

Another way to characterize the incident energy distribution is to cite the median energy.
For the distribution shown in Figure~\ref{fig11} the median energy is 1.36 MeV.
In Figure~\ref{fig12} we show the median energies for all 55 magrols used in this work.
While there is a trend upward in the medians due to the evolving incident energy spectra with time,
the median of the medians  is 1.29 MeV.
Thus the $\sim$0.5-35 MeV rate can also be considered as a rate
of $\sim$1.3 MeV protons.
This characterization of this rate is only appropriate when the energy spectra are
similar to the shape they have in the heliosheath.

\clearpage
\section{Estimation of GCR Background} \label{sec:app2} 
%%%%%%%%%%%%%%%%%%%%%%%%%%%%%%%%%%%%%%%%%%%%%%%%%%%%%%%%%%%%%%%%%
%%%%%%%%%%%%%%%%%%%%%%%%%%%%%%% %%%%%%%%%%%%%%%%%%%%%%%%%%%%%%%%%%%

When V2 crossed the heliopause on day 309 of 2018, the counting rates from the detectors used in this study
gradually dropped from $\sim$25 s$^{-1}$ at the time of the crossing to $\sim$2 s$^{-1}$
at and beyond day 10 of 2019 \citep{StCu19}.
This decline represented the effect of the ACRs escaping into the VLISM.
The residual counting rate of $\sim$2 s$^{-1}$ represents a background due
to galactic cosmic rays. 
To estimate this background rate as a function of time, 
we used a high-energy GCR rate (named PENL) from the HET 2 telescope
and correlated it with the L1 rates during times we believe all the rate is due
to this GCR background source.

\begin{figure}
\begin{center}
\includegraphics*[trim= 0 0.7in 0 1.25in,width=0.5\textwidth]{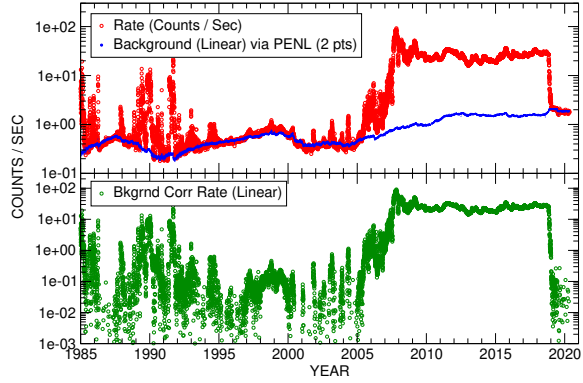}
\end{center}
\caption{(top panel) Counting rate of LET D L1 detector vs. time. 
Also shown as the blue points are the 
estimated GCR background rates based on the PENL rate as described in the text.
(bottom panel) LET D L1 rate corrected for GCR background.
}
%%\label{v2_PENL_vs_LETD_correction_linear_noerror_2point_fit_no_labels}
\label{fig13}
\end{figure}

The correlation for LET D L1 rate is shown in  Figure~\ref{fig13}.
The blue curve represents the background formula that was used for Dbkg in Equation~\ref{eq:2}:
$\textrm{Dbkg} = 3.8\textrm{PENL} - 0.0915$.
LET D was typically in the L1-only command state (configuration 1).
In some cases, denoted as configuration 2, it was put into the L1$\overline{\textrm{L4}}$ command state
just before a magrol and then returned to the configuration 1 state just after the roll ended.
Thus, since this correlation with the PENL rate was only done for configuration 1, for configuration 2  we adjusted the Dbkg value by the ratio of 
the rates obtained during those rolls, L1$\overline{\textrm{L4}}$/L1.

\begin{figure}
\begin{center}
\includegraphics*[trim= 0 0.7in 0 1.25in,width=0.5\textwidth]{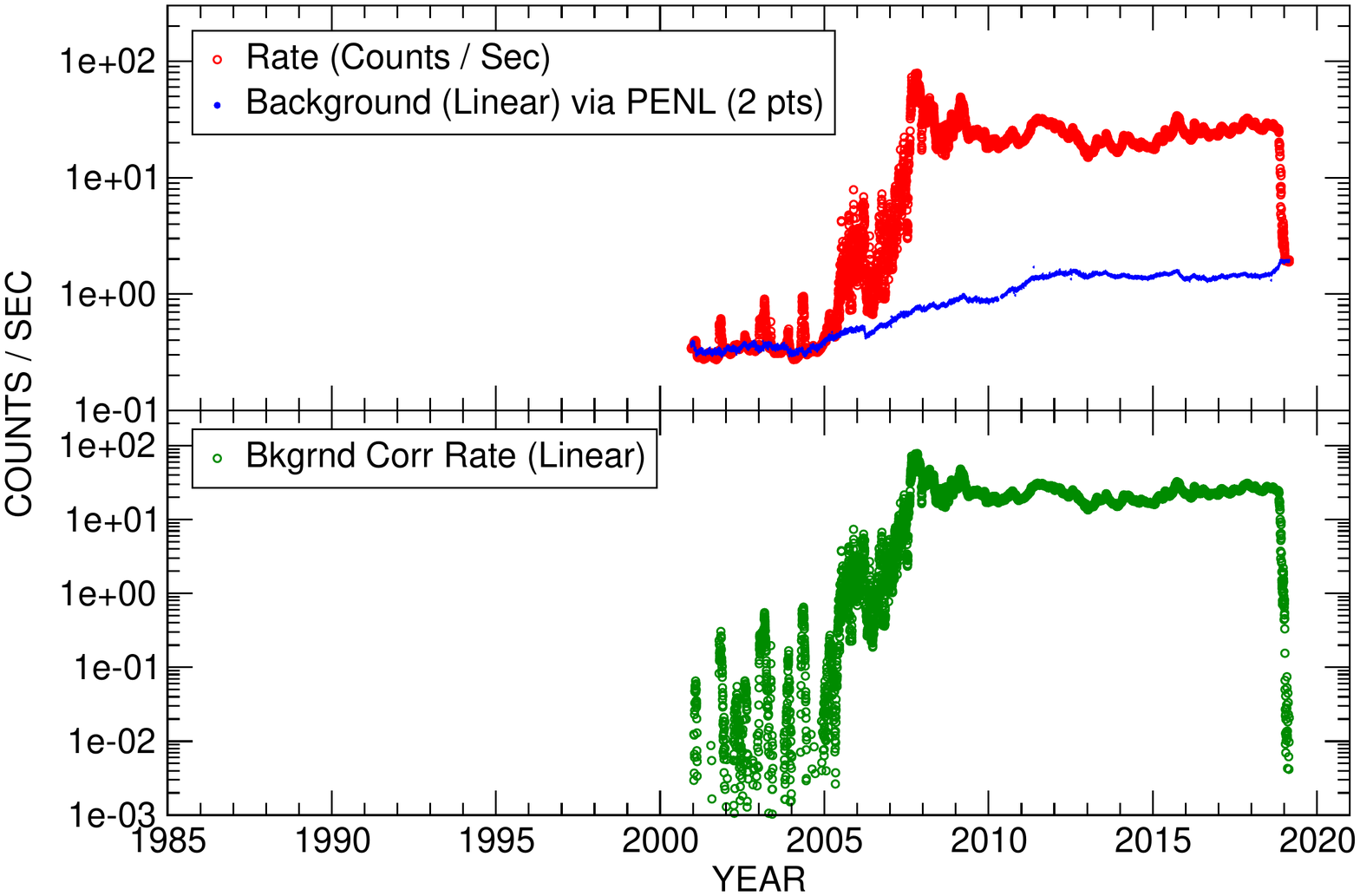}
\end{center}
\caption{Same as Figure~\ref{fig13} except
for the LET A  L1$\overline{\textrm{L4}}$ rate.
}
%%\label{v2_PENL_vs_LETAL1_correction_linear_noerror_2point_fit_no_labels}
\label{fig14}
\end{figure}

\begin{figure}
\begin{center}
\includegraphics*[trim= 0 3.5in 0 1.4in,width=0.5\textwidth]{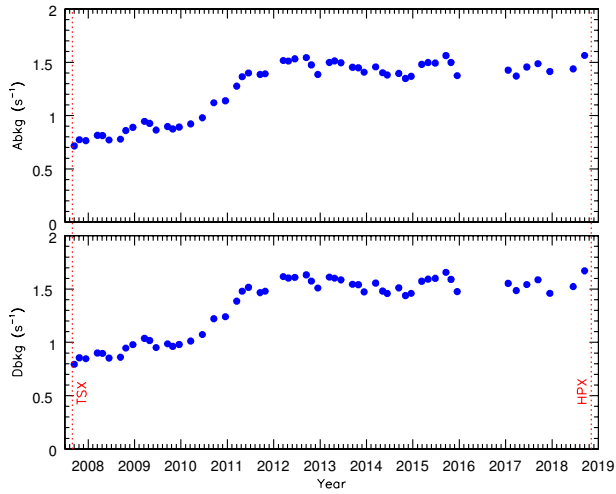}
\end{center}
\caption{
The background counting rates, Abkg and Dbkg in Equations~\ref{eq:1} and \ref{eq:2}, for all 55 magrols in the heliosheath.
}
%%\label{v2_Abkg_Dbkg_time}
\label{fig15}
\end{figure}

The correlation for LET A L1$\overline{\textrm{L4}}$ with the PENL rate is shown in Figure~\ref{fig14}.
LET A was permanently put into the command state L1$\overline{\textrm{L4}}$ in 2000, so only
the time after that is available for the correlation.
The blue curve in the top panel is the result of the equation
$\textrm{Abkg} = 3.6\textrm{PENL} - 0.124$ and this was used in the fits to Equation~\ref{eq:1}.
The background counting rates for all 55 magrols are shown in Figure~\ref{fig15}.

%%%%%%%%%%%%%%%%%%%%%%%%%%%%%%% Figure 13 %%%%%%%%%%%%%%%%%%%%%%%%%%%%%%%%%%%
%%%%%%%%%%%%%%%%%%%%%%%%%%%%%%%%%%%%%%%%%%%%%%%%%%%%%%%%%%%%%%%%%%%%%%%%%%%

%%%%%%%%%%%%%%%%%%%%%%%%%%%%%%% Figure 14 %%%%%%%%%%%%%%%%%%%%%%%%%%%%%%%%%%
%%%%%%%%%%%%%%%%%%%%%%%%%%%%%%%%%%%%%%%%%%%%%%%%%%%%%%%%%%%%%%%%%%%%%%%%%%%%%

%%%%%%%%%%%%%%%%%%%%%%%%%%%%%%%   Figure 15     %%%%%%%%%%%%%%%%%%%%%%%%%%%%%%%%%%%
%%%%%%%%%%%%%%%%%%%%%%%%%%%%%%%%%%%%%%%%%%%%%%%%%%%%%%%%%%%%%%%%%%%%%%%%%%%%%%%%

\clearpage
\section{Tables of Parameters} \label{sec:app3}
%%%%%%%%%%%%%%%%%%%%%%%%%%%%%%%%%%%%%%%%%%%%%%%%%%%%%%%%%%%%%%%%%%%%%%%%%%%%%%%%
%%%%%%%%%%%%%%%%%%%%%%%%%%%%%%% %%%%%%%%%%%%%%%%%%%%%%%%%%%%%%%%%%%

\begin{deluxetable}{lrrrrrrrrrr}
\tablewidth{0pt}
\rotate
\tablecaption{Results from anisotropy fits for all good-fit rolls for 0.5-35 MeV protons}
\label{tab1:results}
\tablecolumns{11}
\tablehead{
\colhead{Roll } & \colhead{Year} & \colhead{Day}  & \colhead{Jo}  & \colhead{$\delta_R$ (true)} & \colhead{$\delta_T$ (true)} & \colhead{$\delta_N$ (true)}  & \colhead{$k_D$} & \colhead{Abkg} & \colhead{Dbkg}  & \colhead{F}
}
\startdata
 40 & 2007 & 256 & 46.43 $\pm$ 0.08 &   1.82e-02 $\pm$ 4.7e-03 &   2.88e-02 $\pm$ 1.9e-03 & -1.98e-02 $\pm$ 1.9e-03 & 1.04630 & 0.7155 & 0.7946 & 3458.3 \\ 
 43 & 2008 &  73 & 45.27 $\pm$ 0.13 &   3.46e-02 $\pm$ 8.0e-03 &   1.87e-02 $\pm$ 3.4e-03 & -2.29e-02 $\pm$ 3.4e-03 & 1.04659 & 0.8150 & 0.8997 & 3617.8 \\ 
 44 & 2008 & 114 & 35.65 $\pm$ 0.06 &   3.45e-02 $\pm$ 5.3e-03 &   1.28e-03 $\pm$ 2.3e-03 & -2.64e-02 $\pm$ 2.3e-03 & 1.04595 & 0.8121 & 0.8966 & 3366.5 \\ 
 45 & 2008 & 165 & 17.14 $\pm$ 0.05 &   7.08e-02 $\pm$ 7.9e-03 &   4.14e-03 $\pm$ 3.3e-03 & -2.86e-02 $\pm$ 3.3e-03 & 1.04358 & 0.7711 & 0.8534 & 2744.2 \\ 
 46 & 2008 & 255 & 19.89 $\pm$ 0.05 &   1.65e-01 $\pm$ 7.3e-03 &   1.30e-01 $\pm$ 3.0e-03 &  2.50e-02 $\pm$ 3.0e-03 & 1.04550 & 0.7785 & 0.8612 & 3077.5 \\ 
 47 & 2008 & 297 & 27.25 $\pm$ 0.07 &   9.89e-02 $\pm$ 6.2e-03 &   2.32e-02 $\pm$ 2.6e-03 & -1.33e-02 $\pm$ 2.5e-03 & 1.04439 & 0.8591 & 0.9462 & 2663.2 \\ 
 48 & 2008 & 353 & 26.45 $\pm$ 0.07 &   4.50e-02 $\pm$ 7.8e-03 &   3.36e-02 $\pm$ 3.2e-03 & -2.34e-02 $\pm$ 3.2e-03 & 1.04674 & 0.8898 & 0.9787 & 3364.4 \\ 
 49 & 2009 &  78 & 42.68 $\pm$ 0.07 &   5.91e-02 $\pm$ 5.0e-03 &   3.09e-02 $\pm$ 2.1e-03 &  2.01e-02 $\pm$ 2.1e-03 & 1.04584 & 0.9459 & 1.0379 & 3170.6 \\ 
 50 & 2009 & 119 & 27.44 $\pm$ 0.07 &   6.75e-02 $\pm$ 6.1e-03 &  -1.30e-02 $\pm$ 2.6e-03 & -1.32e-02 $\pm$ 2.6e-03 & 1.04528 & 0.9267 & 1.0175 & 3184.2 \\ 
 51 & 2009 & 170 & 21.82 $\pm$ 0.06 &   6.83e-02 $\pm$ 6.9e-03 &   9.84e-03 $\pm$ 2.9e-03 & -2.02e-02 $\pm$ 2.9e-03 & 1.04394 & 0.8637 & 0.9511 & 2884.5 \\ 
 52 & 2009 & 260 & 22.83 $\pm$ 0.06 &   7.32e-02 $\pm$ 6.8e-03 &   1.39e-02 $\pm$ 2.8e-03 & -3.19e-02 $\pm$ 2.8e-03 & 1.04473 & 0.8970 & 0.9863 & 3063.2 \\ 
 53 & 2009 & 300 & 22.44 $\pm$ 0.06 &   7.46e-02 $\pm$ 7.0e-03 &  -4.68e-03 $\pm$ 2.8e-03 & -4.20e-02 $\pm$ 2.8e-03 & 1.04428 & 0.8742 & 0.9622 & 2941.1 \\ 
 54 & 2009 & 351 & 20.53 $\pm$ 0.06 &   8.62e-02 $\pm$ 7.3e-03 &   3.86e-03 $\pm$ 3.0e-03 & -3.75e-02 $\pm$ 3.0e-03 & 1.04374 & 0.8913 & 0.9802 & 2863.1 \\ 
 55 & 2010 &  77 & 20.55 $\pm$ 0.06 &   5.31e-02 $\pm$ 7.1e-03 &   9.87e-03 $\pm$ 3.0e-03 & -2.39e-02 $\pm$ 3.0e-03 & 1.04503 & 0.9217 & 1.0123 & 3143.2 \\ 
 56 & 2010 & 169 & 19.68 $\pm$ 0.05 &   7.78e-02 $\pm$ 7.3e-03 &   1.98e-02 $\pm$ 3.1e-03 & -2.26e-02 $\pm$ 3.1e-03 & 1.04335 & 0.9800 & 1.0738 & 2744.3 \\ 
 57 & 2010 & 259 & 23.16 $\pm$ 0.06 &   5.21e-02 $\pm$ 6.8e-03 &   3.53e-02 $\pm$ 2.8e-03 & -2.25e-02 $\pm$ 2.8e-03 & 1.04450 & 1.1197 & 1.2213 & 2935.2 \\ 
 58 & 2010 & 350 & 18.50 $\pm$ 0.05 &   3.99e-02 $\pm$ 7.7e-03 &   2.37e-02 $\pm$ 3.2e-03 & -2.95e-02 $\pm$ 3.2e-03 & 1.04443 & 1.1384 & 1.2411 & 2811.0 \\ 
 59 & 2011 &  74 & 21.98 $\pm$ 0.06 &   2.90e-02 $\pm$ 7.7e-03 &   1.14e-03 $\pm$ 3.3e-03 & -2.49e-02 $\pm$ 3.3e-03 & 1.04930 & 1.2762 & 1.3864 & 3611.1 \\ 
 60 & 2011 & 117 & 26.95 $\pm$ 0.08 &   2.70e-02 $\pm$ 7.1e-03 &   1.35e-02 $\pm$ 3.0e-03 & -1.29e-02 $\pm$ 3.0e-03 & 1.04776 & 1.3646 & 1.4798 & 3056.7 \\ 
 61 & 2011 & 167 & 29.71 $\pm$ 0.07 &   3.71e-02 $\pm$ 5.9e-03 &   2.76e-02 $\pm$ 2.5e-03 & -1.28e-02 $\pm$ 2.5e-03 & 1.04744 & 1.3993 & 1.5164 & 3101.7 \\ 
 62 & 2011 & 258 & 28.49 $\pm$ 0.03 &   3.01e-02 $\pm$ 2.2e-03 &   1.18e-02 $\pm$ 2.0e-03 & -2.33e-02 $\pm$ 2.0e-03 & 1.02955 & 1.3857 & 1.4669 & 3216.9 \\ 
 63 & 2011 & 298 & 26.99 $\pm$ 0.03 &   3.36e-02 $\pm$ 2.2e-03 &   1.62e-02 $\pm$ 2.1e-03 & -2.01e-02 $\pm$ 2.1e-03 & 1.02852 & 1.3920 & 1.4803 & 2898.9 \\ 
 64 & 2012 &  75 & 25.34 $\pm$ 0.03 &   3.66e-02 $\pm$ 2.3e-03 &   7.58e-03 $\pm$ 2.2e-03 & -1.78e-02 $\pm$ 2.2e-03 & 1.02922 & 1.5160 & 1.6167 & 3131.6 \\ 
 65 & 2012 & 116 & 22.46 $\pm$ 0.02 &   2.72e-02 $\pm$ 2.4e-03 &   7.36e-03 $\pm$ 2.3e-03 & -1.27e-02 $\pm$ 2.3e-03 & 1.02889 & 1.5112 & 1.6037 & 3018.5 \\ 
 66 & 2012 & 167 & 20.24 $\pm$ 0.02 &   1.17e-02 $\pm$ 2.6e-03 &   2.59e-02 $\pm$ 2.4e-03 & -3.37e-03 $\pm$ 2.4e-03 & 1.03011 & 1.5318 & 1.6092 & 3579.1 
\enddata
\tablecomments{Roll number created for purposes of this paper; units of Jo are s$^{-1}$; date information is for start of roll; true refers to direction of flow in the Sun frame  and corrected for wide 
field of view of telescope; units of Abkg and Dbkg are s$^{-1}$; F is C-G factor in units of km s$^{-1}$ with estimated uncertainty of 204 km s$^{-1}$.}
\end{deluxetable}

\begin{deluxetable}{lrrrrrrrrrr}
\tablewidth{0pt}
\rotate
\tablecaption{Results from anisotropy fits for all good-fit rolls for 0.5-35 MeV protons (continued)}
\label{tab2:results}
\tablecolumns{11}
\tablehead{
\colhead{Roll } & \colhead{Year} & \colhead{Day}  & \colhead{Jo}  & \colhead{$\delta_R$ (true)} & \colhead{$\delta_T$ (true)} & \colhead{$\delta_N$ (true)}  & \colhead{$k_D$} & \colhead{Abkg} & \colhead{Dbkg} &  \colhead{F}
}
\startdata
 67 & 2012 & 257 & 20.37 $\pm$ 0.02 &   2.66e-02 $\pm$ 2.6e-03 &   5.02e-02 $\pm$ 2.4e-03 & -1.82e-02 $\pm$ 2.4e-03 & 1.02975 & 1.5438 & 1.6326 & 3263.7 \\ 
 68 & 2012 & 297 & 17.71 $\pm$ 0.03 &   2.83e-02 $\pm$ 2.8e-03 &   1.08e-02 $\pm$ 2.6e-03 & -1.40e-02 $\pm$ 2.6e-03 & 1.02903 & 1.4750 & 1.5744 & 3005.4 \\ 
 69 & 2012 & 348 & 15.26 $\pm$ 0.05 &   3.09e-02 $\pm$ 7.3e-03 &   2.88e-02 $\pm$ 7.1e-03 & -1.16e-02 $\pm$ 6.8e-03 & 1.02859 & 1.3854 & 1.5095 & 2932.1 \\ 
 70 & 2013 &  73 & 14.91 $\pm$ 0.03 &   3.06e-02 $\pm$ 3.1e-03 &   8.03e-02 $\pm$ 3.0e-03 & -8.37e-03 $\pm$ 3.0e-03 & 1.02806 & 1.4982 & 1.6124 & 2907.7 \\ 
 72 & 2013 & 164 & 18.16 $\pm$ 0.03 &   2.46e-02 $\pm$ 2.7e-03 &   3.25e-03 $\pm$ 2.6e-03 & -1.94e-02 $\pm$ 2.6e-03 & 1.02910 & 1.4958 & 1.5865 & 3101.0 \\ 
 73 & 2013 & 255 & 17.99 $\pm$ 0.03 &   3.35e-02 $\pm$ 2.8e-03 &   9.76e-03 $\pm$ 2.6e-03 & -2.31e-02 $\pm$ 2.6e-03 & 1.02915 & 1.4529 & 1.5449 & 3077.4 \\ 
 75 & 2013 & 346 & 15.14 $\pm$ 0.02 &   2.75e-02 $\pm$ 3.0e-03 &   2.32e-02 $\pm$ 2.8e-03 & -7.21e-03 $\pm$ 2.8e-03 & 1.02944 & 1.4068 & 1.4732 & 3227.3 \\ 
 76 & 2014 &  72 & 19.77 $\pm$ 0.02 &   3.23e-02 $\pm$ 2.6e-03 &   1.42e-02 $\pm$ 2.5e-03 & -1.83e-02 $\pm$ 2.5e-03 & 1.03095 & 1.4563 & 1.5556 & 3565.2 \\ 
 77 & 2014 & 127 & 19.01 $\pm$ 0.02 &   4.28e-02 $\pm$ 2.7e-03 &   2.10e-02 $\pm$ 2.5e-03 & -1.57e-02 $\pm$ 2.5e-03 & 1.03005 & 1.4013 & 1.4807 & 3125.9 \\ 
 78 & 2014 & 163 & 17.72 $\pm$ 0.03 &   3.70e-02 $\pm$ 2.8e-03 &   3.61e-02 $\pm$ 2.6e-03 & -1.88e-02 $\pm$ 2.6e-03 & 1.02945 & 1.3808 & 1.4583 & 3189.5 \\ 
 79 & 2014 & 254 & 17.95 $\pm$ 0.05 &   3.05e-02 $\pm$ 8.0e-03 &   1.04e-02 $\pm$ 3.3e-03 & -1.96e-02 $\pm$ 3.3e-03 & 1.04973 & 1.3948 & 1.5117 & 3343.9 \\ 
 80 & 2014 & 307 & 16.67 $\pm$ 0.03 &   1.53e-02 $\pm$ 2.9e-03 &   2.21e-02 $\pm$ 2.7e-03 & -1.49e-02 $\pm$ 2.7e-03 & 1.02929 & 1.3487 & 1.4392 & 3191.7 \\ 
 81 & 2014 & 352 & 16.41 $\pm$ 0.03 &   1.36e-02 $\pm$ 2.9e-03 &   1.22e-02 $\pm$ 2.7e-03 & -1.42e-02 $\pm$ 2.7e-03 & 1.02956 & 1.3676 & 1.4600 & 3474.0 \\ 
 82 & 2015 &  70 & 20.71 $\pm$ 0.02 &   1.71e-02 $\pm$ 2.5e-03 &   1.49e-02 $\pm$ 2.4e-03 & -1.43e-02 $\pm$ 2.4e-03 & 1.02959 & 1.4799 & 1.5735 & 3272.8 \\ 
 83 & 2015 & 119 & 19.05 $\pm$ 0.02 &   1.00e-02 $\pm$ 2.7e-03 &   2.07e-02 $\pm$ 2.5e-03 & -1.19e-02 $\pm$ 2.5e-03 & 1.02985 & 1.4970 & 1.5927 & 3393.0 \\ 
 84 & 2015 & 176 & 21.54 $\pm$ 0.02 &   2.06e-02 $\pm$ 2.5e-03 &   2.10e-02 $\pm$ 2.3e-03 & -1.49e-02 $\pm$ 2.3e-03 & 1.02957 & 1.4918 & 1.6001 & 3267.9 \\ 
 85 & 2015 & 260 & 29.81 $\pm$ 0.03 &   4.11e-02 $\pm$ 2.1e-03 &   1.20e-02 $\pm$ 2.0e-03 & -4.67e-03 $\pm$ 2.0e-03 & 1.02962 & 1.5620 & 1.6571 & 3233.4 \\ 
 86 & 2015 & 300 & 27.86 $\pm$ 0.03 &   5.50e-02 $\pm$ 2.2e-03 &   6.62e-02 $\pm$ 2.0e-03 &  1.64e-02 $\pm$ 2.0e-03 & 1.02938 & 1.4984 & 1.5914 & 2985.0 \\ 
 87 & 2015 & 349 & 23.10 $\pm$ 0.02 &   3.66e-02 $\pm$ 2.4e-03 &   1.54e-02 $\pm$ 2.2e-03 & -1.54e-02 $\pm$ 2.3e-03 & 1.02996 & 1.3743 & 1.4762 & 3111.2 \\ 
 88 & 2017 &  20 & 23.58 $\pm$ 0.08 &   3.00e-02 $\pm$ 7.9e-03 &   4.00e-02 $\pm$ 7.5e-03 &  2.13e-03 $\pm$ 7.6e-03 & 1.02993 & 1.4260 & 1.5529 & 3282.1 \\ 
 89 & 2017 &  83 & 23.38 $\pm$ 0.06 &   2.10e-02 $\pm$ 5.3e-03 &   3.05e-02 $\pm$ 5.1e-03 &  2.82e-03 $\pm$ 5.1e-03 & 1.02948 & 1.3705 & 1.4855 & 3180.0 \\ 
 90 & 2017 & 166 & 25.29 $\pm$ 0.06 &   1.64e-02 $\pm$ 5.1e-03 &   1.65e-02 $\pm$ 4.8e-03 & -9.61e-03 $\pm$ 4.8e-03 & 1.02993 & 1.4554 & 1.5423 & 3401.7 \\ 
 91 & 2017 & 254 & 24.83 $\pm$ 0.06 &   3.46e-03 $\pm$ 5.2e-03 &   3.73e-02 $\pm$ 4.9e-03 & -2.97e-03 $\pm$ 4.9e-03 & 1.03092 & 1.4868 & 1.5866 & 3844.9 \\ 
 92 & 2017 & 348 & 28.33 $\pm$ 0.08 &   1.86e-02 $\pm$ 5.3e-03 &  -1.68e-03 $\pm$ 4.9e-03 & -1.26e-02 $\pm$ 5.1e-03 & 1.02998 & 1.4135 & 1.4601 & 3340.9 \\ 
 93 & 2018 & 166 & 27.07 $\pm$ 0.07 &   1.83e-02 $\pm$ 5.0e-03 &   2.42e-02 $\pm$ 4.7e-03 &  7.25e-03 $\pm$ 4.7e-03 & 1.03052 & 1.4375 & 1.5228 & 3601.1 \\ 
 94 & 2018 & 256 & 25.82 $\pm$ 0.07 &   1.09e-02 $\pm$ 5.1e-03 &   6.14e-03 $\pm$ 4.8e-03 &  2.97e-03 $\pm$ 4.8e-03 & 1.03066 & 1.5634 & 1.6714 & 3853.0 
\enddata
\end{deluxetable}

\clearpage

%%\bibliography{ms}

\begin{thebibliography}{}
\expandafter\ifx\csname natexlab\endcsname\relax\def\natexlab#1{#1}\fi
\providecommand{\url}[1]{\href{#1}{#1}}
\providecommand{\dodoi}[1]{doi:~\href{http://doi.org/#1}{\nolinkurl{#1}}}
\providecommand{\doeprint}[1]{\href{http://ascl.net/#1}{\nolinkurl{http://ascl.net/#1}}}
\providecommand{\doarXiv}[1]{\href{https://arxiv.org/abs/#1}{\nolinkurl{https://arxiv.org/abs/#1}}}

\bibitem[{{Breneman}(1985)}]{Br85}
{Breneman}, H.~H. 1985, PhD thesis, California Institute of Technology,
  Pasadena.

\bibitem[{{Bridge} {et~al.}(1977){Bridge}, {Belcher}, {Butler}, {Lazarus},
  {Mavretic}, {Sullivan}, {Siscoe}, \& {Vasyliunas}}]{BrBe77}
{Bridge}, H.~S., {Belcher}, J.~W., {Butler}, R.~J., {et~al.} 1977, \ssr, 21,
  259, \dodoi{10.1007/BF00211542}

\bibitem[{{Burlaga} \& {Ness}(2012)}]{BuNe12}
{Burlaga}, L.~F., \& {Ness}, N.~F. 2012, \apj, 749, 13,
  \dodoi{10.1088/0004-637X/749/1/13}

\bibitem[{{Cook}(1981)}]{Co81}
{Cook}, W.~R., I. 1981, PhD thesis, California Institute of Technology,
  Pasadena.

\bibitem[{{Cook} {et~al.}(1984){Cook}, {Stone}, \& {Vogt}}]{CoSt84}
{Cook}, W.~R., {Stone}, E.~C., \& {Vogt}, R.~E. 1984, \apj, 279, 827,
  \dodoi{10.1086/161953}

\bibitem[{{Cummings} {et~al.}(2019){Cummings}, {Stone}, {Heikkila}, {Lal}, \&
  {Richardson}}]{CuSt19}
{Cummings}, A., {Stone}, E., {Heikkila}, B.~C., {Lal}, N., \& {Richardson}, J.
  2019, in International Cosmic Ray Conference, Vol.~36, 36th International
  Cosmic Ray Conference (ICRC2019), 1071

\bibitem[{{Cummings} {et~al.}(2016){Cummings}, {Stone}, {Heikkila}, {Lal},
  {Webber}, {J{\'o}hannesson}, {Moskalenko}, {Orlando}, \& {Porter}}]{CuSt16}
{Cummings}, A.~C., {Stone}, E.~C., {Heikkila}, B.~C., {et~al.} 2016, \apj, 831,
  18, \dodoi{10.3847/0004-637X/831/1/18}

\bibitem[{{Decker} {et~al.}(2012){Decker}, {Krimigis}, {Roelof}, \&
  {Hill}}]{DeKr12}
{Decker}, R.~B., {Krimigis}, S.~M., {Roelof}, E.~C., \& {Hill}, M.~E. 2012,
  \nat, 489, 124, \dodoi{10.1038/nature11441}

\bibitem[{{Drake} {et~al.}(2017){Drake}, {Swisdak}, {Opher}, \&
  {Richardson}}]{DrSw17}
{Drake}, J.~F., {Swisdak}, M., {Opher}, M., \& {Richardson}, J.~D. 2017, \apj,
  837, 159, \dodoi{10.3847/1538-4357/aa6304}

\bibitem[{{Forman}(1970)}]{Fo70}
{Forman}, M.~A. 1970, Planet. Space Sci., 18, 25,
  \dodoi{10.1016/0032-0633(70)90064-4}

\bibitem[{{Fr{\"a}nz} \& {Harper}(2002)}]{FrHa02}
{Fr{\"a}nz}, M., \& {Harper}, D. 2002, \planss, 50, 217,
  \dodoi{10.1016/S0032-0633(01)00119-2}

\bibitem[{{Kane} {et~al.}(1998){Kane}, {Decker}, {Mauk}, \&
  {Krimigis}}]{KaDe98}
{Kane}, M., {Decker}, R.~B., {Mauk}, B.~H., \& {Krimigis}, S.~M. 1998, \jgr,
  103, 267, \dodoi{10.1029/97JA02776}

\bibitem[{{Krimigis} {et~al.}(1977){Krimigis}, {Armstrong}, {Axford},
  {Bostrom}, {Fan}, {Gloeckler}, \& {Lanzerotti}}]{KrAr77}
{Krimigis}, S.~M., {Armstrong}, T.~P., {Axford}, W.~I., {et~al.} 1977, \ssr,
  21, 329, \dodoi{10.1007/BF00211545}

\bibitem[{{Krimigis} {et~al.}(2013){Krimigis}, {Decker}, {Roelof}, {Hill},
  {Armstrong}, {Gloeckler}, {Hamilton}, \& {Lanzerotti}}]{KrDe13}
{Krimigis}, S.~M., {Decker}, R.~B., {Roelof}, E.~C., {et~al.} 2013, Science,
  341, 144, \dodoi{10.1126/science.1235721}

\bibitem[{{Krimigis} {et~al.}(2011){Krimigis}, {Roelof}, {Decker}, \&
  {Hill}}]{KrRo11}
{Krimigis}, S.~M., {Roelof}, E.~C., {Decker}, R.~B., \& {Hill}, M.~E. 2011,
  \nat, 474, 359, \dodoi{10.1038/nature10115}

\bibitem[{{Opher} {et~al.}(2012){Opher}, {Drake}, {Velli}, {Decker}, \&
  {Toth}}]{OpDr12}
{Opher}, M., {Drake}, J.~F., {Velli}, M., {Decker}, R.~B., \& {Toth}, G. 2012,
  \apj, 751, 80, \dodoi{10.1088/0004-637X/751/2/80}

\bibitem[{{Pogorelov} {et~al.}(2012){Pogorelov}, {Borovikov}, {Zank},
  {Burlaga}, {Decker}, \& {Stone}}]{PoBo12}
{Pogorelov}, N.~V., {Borovikov}, S.~N., {Zank}, G.~P., {et~al.} 2012, \apjl,
  750, L4, \dodoi{10.1088/2041-8205/750/1/L4}

\bibitem[{{Pogorelov} {et~al.}(2017){Pogorelov}, {Heerikhuisen},
  {Roytershteyn}, {Burlaga}, {Gurnett}, \& {Kurth}}]{PoHe17}
{Pogorelov}, N.~V., {Heerikhuisen}, J., {Roytershteyn}, V., {et~al.} 2017,
  \apj, 845, 9, \dodoi{10.3847/1538-4357/aa7d4f}

\bibitem[{{Press} {et~al.}(1992){Press}, {Teukolsky}, {Vetterling}, \&
  {Flannery}}]{PrTe92}
{Press}, W.~H., {Teukolsky}, S.~A., {Vetterling}, W.~T., \& {Flannery}, B.~P.
  1992, {Numerical recipes in C. The art of scientific computing}

\bibitem[{{Richardson} {et~al.}(2020){Richardson}, {Belcher}, {Burlaga},
  {Cummings}, {Decker}, {Opher}, \& {Stone}}]{RiBe20}
{Richardson}, J.~D., {Belcher}, J.~W., {Burlaga}, L.~F., {et~al.} 2020, in IOP
  Conference Series, Vol. tbd, From the Sun's Atmosphere to the Edge of the
  Galaxy: A Story of Connections: 19th Annual International Astrophysics
  Conference, ed. G.~P. {Zank}, tbd, \dodoi{10.1088/1742-6596/1620/1/012016,
  2020}

\bibitem[{{Richardson} {et~al.}(2018){Richardson}, {Belcher}, {Cummings},
  {Decker}, \& {Stone}}]{RiBe18}
{Richardson}, J.~D., {Belcher}, J.~W., {Cummings}, A.~C., {Decker}, R., \&
  {Stone}, E.~C. 2018, in Journal of Physics Conference Series, Vol. 1100,
  Journal of Physics Conference Series, 012019,
  \dodoi{10.1088/1742-6596/1100/1/012019}

\bibitem[{{Richardson} {et~al.}(2013){Richardson}, {Burlaga}, {Decker},
  {Drake}, {Ness}, \& {Opher}}]{RiBu13}
{Richardson}, J.~D., {Burlaga}, L.~F., {Decker}, R.~B., {et~al.} 2013, \apjl,
  762, L14, \dodoi{10.1088/2041-8205/762/1/L14}

\bibitem[{{Richardson} \& {Decker}(2014)}]{RiDe14}
{Richardson}, J.~D., \& {Decker}, R.~B. 2014, Astrophys. J., 792, 126,
  \dodoi{10.1088/0004-637X/792/2/126}

\bibitem[{{Stone} {et~al.}(2017){Stone}, {Cummings}, {Heikkila}, {Lal}, \&
  {Webber}}]{StCu17}
{Stone}, E., {Cummings}, A.~C., {Heikkila}, B.~C., {Lal}, N., \& {Webber},
  W.~R. 2017, International Cosmic Ray Conference, 301, 57

\bibitem[{{Stone} \& {Cummings}(2003)}]{StCu03}
{Stone}, E.~C., \& {Cummings}, A.~C. 2003, in International Cosmic Ray
  Conference, Vol.~7, International Cosmic Ray Conference, 3781

\bibitem[{{Stone} \& {Cummings}(2011)}]{StCu11}
{Stone}, E.~C., \& {Cummings}, A.~C. 2011, International Cosmic Ray Conference,
  12, 29, \dodoi{10.7529/ICRC2011/V12/I06}

\bibitem[{{Stone} {et~al.}(2019){Stone}, {Cummings}, {Heikkila}, \&
  {Lal}}]{StCu19}
{Stone}, E.~C., {Cummings}, A.~C., {Heikkila}, B.~C., \& {Lal}, N. 2019, Nature
  Astronomy, 3, 1013, \dodoi{10.1038/s41550-019-0928-3}

\bibitem[{Stone {et~al.}(1977)Stone, Vogt, McDonald, Teegarden, Trainor,
  Jokipii, \& Webber}]{StVo77}
Stone, E.~C., Vogt, R.~E., McDonald, F.~B., {et~al.} 1977, Space Sci. Rev., 21,
  355

\bibitem[{{Sullivan}(1971)}]{Su71}
{Sullivan}, J.~D. 1971, Nuclear Instruments and Methods, 95, 5,
  \dodoi{10.1016/0029-554X(71)90033-4}

\end{thebibliography}
\bibliographystyle{aasjournal}



\end{document}